\documentclass{article}%
\usepackage{amssymb}
\usepackage{amsmath}%
\usepackage{amsfonts}%
\usepackage{graphicx}
\providecommand{\U}[1]{\protect\rule{.1in}{.1in}}

\begin{document}

\title{An exact solution of the time-dependent Schr\"{o}dinger equation with a
rectangular potential for real and imaginary time }
\author{Victor F. Los* and Mykola "Nicholas" V. Los**
\and *Institute for Magnetism, Nat. Acad. Sci. of Ukraine,
\and 36-b Vernadsky Blvd., Kiev 142, Ukraine
\and **Luxoft Eastern Europe
\and 14 Vasilkovskaya Str., B, Business Center STEND,
\and Kiev 03040, Ukraine}

\begin{abstract}
A propagator for the one dimensional time-dependent Schr\"{o}dinger equation
with an asymmetric rectangular potential is obtained using the
multiple-scattering theory approach. It allows for the consideration of the
reflection and transmission processes as the particle scattering at the
potential jump (in contrast to the conventional wave-like picture) and for
accounting for the nonclassical counterintuitive contribution of the
backward-moving component of the wave packet attributed to the particle. This
propagator completely resolves the corresponding time-dependent
Schr\"{o}dinger equation (defines the wave function $\psi(x,t)$) and allows
for considering the quantum mechanical effects of a particle reflection from
the potential downward step/well and a particle tunneling through the
potential barrier as a function of time. These results are related to
fundamental issues such as measuring time in quantum mechanics (tunneling
time, time of arrival, dwell time). For imaginary time, which represents an
inverse temperature ($t\rightarrow-i\hbar/k_{B}T$), the obtained propagator is
equivalent to the density matrix for a particle that is in a heat bath and is
subject to a rectangular potential. This density matrix provides information
on the particles' density in the different spatial areas relative to the
potential location and on the quantum coherences of the different particle
spatial states. If one shifts to imaginary time ($t\rightarrow-it$), the
matrix element of the calculated propagator in the spatial basis provides a
solution to the diffusion-like equation with a rectangular potential. The
obtained exact results are presented as the integrals from elementary
functions and thus allow for a numerical visualization of the probability
density $\left\vert \psi(x,t)\right\vert ^{2}$, the density matrix and the
solution of the diffusion-like equation. The results obtained may also be
useful for spintronics applications due to the fact that the asymmetric
(spin-dependent) rectangular potential can model the potential profile in
layered magnetic nanostructures..


\end{abstract}
\maketitle

\section{Introduction}

We start with the one-dimensional Schr\"{o}dinger equation for a particle of
mass $m$ subject to potential $V(x)$
\begin{equation}
i\hbar\frac{\partial\psi(x;t)}{\partial t}=H\psi(x;t),\label{0}%
\end{equation}
where $H$ is a self-adjoint operator%
\begin{equation}
H=-\frac{\hbar^{2}}{2m}\frac{\partial^{2}}{\partial x^{2}}+V(x).\label{0a}%
\end{equation}
A solution to this equation can generally be presented as%
\begin{equation}
\psi(x;t)=\int<x|K(t)|x^{\prime}>\psi(x^{\prime};0)dx^{\prime},\label{0b}%
\end{equation}
where $K(t)=\exp(-iHt/\hbar)$ is the propagator (Green's function) for
equation (\ref{0}) in operator form and $<x|K(t)|x^{\prime}>$ is its matrix
element in $x$-representation. Thus, the knowledge of the propagator provides
the complete solution to the equation (\ref{0}) at the given initial value
$\psi(x^{\prime};0)$. If the initial value is of the form $\psi(x^{\prime
};0)=\delta(x^{\prime}-x^{\prime\prime})$, the solution (\ref{0b}) reduces to
the Green function matrix element
\begin{equation}
\psi(x,x^{\prime\prime};t)=<x|K(t)|x^{\prime\prime}>.\label{0c}%
\end{equation}
\qquad

Equation (\ref{0}) with an imaginary time variable is also relevant to other
physical situations. If we make the substitutions $t\rightarrow-i\hbar\beta$
($\beta=1/k_{B}T$) and $\psi(x,x^{\prime};-i\hbar\beta)\rightarrow
\rho(x,x^{\prime};\beta)$, Eq. (\ref{0c}) represents the matrix element
$\rho(x,x^{\prime};\beta)=<x|\exp(-\beta H)|x^{\prime}>$ of the density
operator $\rho(\beta)=\exp(-\beta H)$, which satisfies the Bloch equation (in
the $x$-representation)
\begin{equation}
\frac{\partial\rho(x,x^{\prime};\beta)}{\partial\beta}=-H\rho(x,x^{\prime
};\beta),\label{0d}%
\end{equation}
with initial condition $\rho(x,x^{\prime};\beta=0)=\delta(x-x^{\prime})$ and
where the operator $H$ (\ref{0a}) in (\ref{0d}) is applied only to the $x $
variable of the density matrix.

If we make the substitutions $t\rightarrow-it$, $\hbar\rightarrow2mD$,
$V(x)/2mD\rightarrow\overline{V}(x)$ and $\psi(x;-it)\rightarrow Q(x;t)$, Eq.
(\ref{0}) represents the inhomogeneous diffusion-like equation (with the
diffusion coefficient $D$)
\begin{equation}
\frac{\partial Q(x;t)}{\partial t}=D\frac{\partial^{2}Q(x;t)}{\partial x^{2}%
}-\overline{V}(x)Q(x;t).\label{0e}%
\end{equation}
The solution to Eq. (\ref{0e}) at the initial condition $Q(x;0)=\delta
(x-x_{0})$ is given by (\ref{0c})%
\begin{equation}
Q(x,x_{0};t)=<x|\exp(-Ht/2mD)|x_{0}>,\label{0f}%
\end{equation}
\qquad where $H$ is defined by (\ref{0a}) with $\hbar\rightarrow2mD$.

We see that, in any case, the problem is to find a propagator of the type
$<x|\exp(-\alpha H)|x^{\prime}>$ with different $\alpha$ for the considered
parabolic differential equations.

A rectangular potential is the simplest one allowing for the study of some
striking quantum mechanical effects, such as particle reflection from a
potential step/well and transmission through a potential barrier. These
phenomena are less surprising when we think of a wave being, e.g., reflected
from a downward potential step, though they are more surprising from the
particle point of view. They easily follow from the standard textbook
stationary analysis, which reduces to substituting a plane wave of energy $E$
for the wave packet and solving the stationary Schr\"{o}dinger equation.
However, in this case, there are no real transport phenomena, i.e. in the
absence of the energy dispersion ($\Delta E=0$) the transmission time through
or the time of arrival (TOA) to the potential jumps is indefinite ($\Delta
t\backsim\hbar/\Delta E$). It is interesting to verify the mentioned
non-classical phenomena by considering the time-dependent picture of these
processes in a realistic situation, when a particle, originally localized
outside the potential well/barrier, moves towards the potential and
experiences scattering at the potential jumps. In order to do this, the
corresponding time-dependent Schr\"{o}dinger equation needs to be solved. This
problem is much more involved even in the one-dimensional case in comparison
to the conventional stationary case.

In particular, there is one striking and classically forbidden
counterintuitive (and often overlooked) effect even in the process of the
simplest 1D time-dependent scattering by the mentioned potentials. A wave
packet representing an ensemble of particles, confined initially (at $t=t_{0}
$), say, somewhere to the region $x<0$, consists of both positive and negative
momentum components due to the fact that a particle cannot be completely
localized at $x<0$ if the wave packet contains only $p>0$ components. One
would then expect that only particles with positive momenta $p$ may arrive at
positive positions $x>0$ at $t>t_{0}$. However, the wave packet's negative
momentum components (restricted to a half line in momentum space) are
necessarily different from zero in the whole $x$ space ($-\infty\div\infty$),
representing the particles' presence at $x>0$ at initial moment of time
$t_{0}$, and, therefore, may contribute, for example, to the distribution of
the particles' time of arrival (TOA) to $x>0$ \cite{Muga Sources 2001,Baute
2002}. It is worth noting that the contribution of the backward-moving
(negative momentum) components in the initial-value problem is in some sense
equivalent to the contribution of the negative energy (evanescent) components
in the source solution \cite{Muga Sources 2001}. Thus, the correct treatment
of some aspects of the kinetics of the wave packet (even in the 1D case and
even for a "free" motion) becomes a nontrivial problem and is closely related
to the fundamental problem of measuring time in quantum mechanics, such as
TOA, the dwell time, and tunneling time.

In addition, the time-dependent aspects of reflection from and transmission
through the potential step/barrier/well have recently acquired relevance not
only in view of renewed interest in the fundamental problems of measuring time
in quantum mechanics (see \cite{Time in QM}), but also due to important
practical applications in the newly emerged fields of nanoscience and
nanotechnology. Rectangular (asymmetric spin-dependent) potential
barriers/wells may often satisfactorily approximate the one-dimensional
potential profiles in layered magnetic nanostructures (with sharp interfaces).
In such nanostructures, the giant magnetoresistance (GMR) \cite{GMR} and
tunneling magnetoresistance (TMR) \cite{TMR} effects occur.

The calculation of the propagator $<x|\exp(-\alpha H)|x^{\prime}>$ is
conveniently related to the path-integral method (see, e.g. \cite{Feynman} and
\cite{Kac}). The list of the exact solutions for this propagator is very
short. For example, there is an exact solution for the spacetime propagator
$<x|\exp(-iHt/\hbar)|x^{\prime}>$ of the Schr\"{o}dinger equation in the
one-dimensional square barrier case obtained in \cite{Barut 1988}, but this
solution is very complicated, implicit and not easy to analyze (see also
\cite{Schulman 1982, Carvalho 1993, Yearsley 2008}).

Recently, we have suggested a simple method for the calculation of the
spacetime propagator \cite{Los 2010,Los 2011,Los 2012}, which exactly resolves
the time-dependent Schr\"{o}dinger equation with a rectangular potential in
terms of integrals of elementary functions. This method is an alternative to
the commonly used path-integral approach to the mentioned problems and based
on the energy integration of the spectral density matrix (discontinuity of the
energy-dependent Green function across the real energy axis). The
energy-dependent Green function is then easily obtained for the
step/barrier/well potentials with multiple-scattering theory (MST) using the
effective energy-dependent potentials found in \cite{Los 2010}, which are
responsible for reflection from and transmission through the potential step.
These potentials, which are defined via the different particle velocities from
both sides of the potential steps making up the step/barrier/well potentials,
allow for the consideration of the reflection and transmission processes as
particle scattering at the potential jumps in contrast to the conventional
wave-like picture. An important advantage of our approach is that the negative
energy (evanescent states) contribution to the propagator cancels out due to
the natural decomposition of the propagator into forward- and backward-moving
components. This is an essential result because accounting for both of these
components (which should generally be done) often leads to a rather
complicated consideration of the evanescent states with $E<0$ (see \cite{Muga
2000}).

In this paper, we provide an exact solution to Eq. (\ref{0}) for real and
imaginary times using our approach \cite{Los 2010,Los 2011,Los 2012} to the
calculation of the spacetime propagator for a general asymmetric rectangular
potential. In Sec. 2, we outline our MST\ approach to the calculation of the
propagator for the time-dependent Schr\"{o}dinger equation and present its
explicit form. In Sec. 3, we consider a system in a heat bath, as is the case,
e.g., for electrons in nanostructures. The equilibrium system's
characteristics can then be calculated knowing its density matrix
$\rho(x,x^{\prime};\beta)=<x|\exp(-\beta H)|x^{\prime}>$. We present in this
section an exact solution for the density matrix of a particle in an
asymmetric (spin-dependent) one-dimensional rectangular potential and discuss
its properties with the help of numerical evaluation of the corresponding
integrals of elementary functions. In accordance with the above discussion of
Eq. (\ref{0}), the obtained solution for the spacetime propagator may be also
used for finding the solution to the diffusion-like equation (\ref{0}) through
the appropriate change of the equation parameters. This case is discussed in
Sec. 4 and the summary of the results is given in Sec. 5.

\section{Multiple scattering calculation of a spacetime propagator for the
Schr\"{o}dinger equation}

We start by considering a particle (electron) of mass $m$ in the following
general asymmetric one-dimensional rectangular potential of width $d$ placed
in the interval $0<x<d$%
\begin{equation}
V(x)=[\theta(x)-\theta(x-d)]U+\theta(x-d)\Delta,\label{1}%
\end{equation}
where $\theta(x)$ is the Heaviside step function, and the potential parameters
$U$ and $\Delta$ can acquire positive as well as negative values (for
$\Delta=U$, $V(x)$ reduces to the step potential). As an application, we can
model by the potential (\ref{1}) a spin-dependent potential profile of a
threelayer made of a nonmagnetic spacer (metallic or insulator) sandwiched
between two magnetic (infinite) layers, and the asymmetricity
(spin-dependence) of the potential (\ref{1}) is defined by the parameter
$\Delta$. The particle wave vectors in different spatial areas (layers) are
defined as%

\begin{align}
k_{<}^{0}(E)  & =k(E),k(E)=\sqrt{\frac{2m}{\hbar^{2}}E},x<0,\nonumber\\
k_{>}^{0}(E)  & =k_{<}^{d}(E)=k_{u}(E),k_{u}(E)=\sqrt{\frac{2m}{\hbar^{2}%
}(E-U)},0<x<d,\nonumber\\
k_{>}^{d}(E)  & =k_{\Delta}(E),k_{\Delta}(E)=\sqrt{\frac{2m}{\hbar^{2}%
}(E-\Delta)},x>d.\label{2}%
\end{align}
In the case of three-dimensional sandwiches, $k_{>(<)}^{0}(E)$ and
$k_{>(<)}^{d}(E)$ are the perpendicular-to-interface components of the wave
vector $\mathbf{k}$ of a particle arriving at the interfaces (located at $x=0
$ and $x=d$) from the right ($>$) or from the left ($<$).

The wave function of a single particle moving in perturbing potential $V(x)$
is given by Eq. (\ref{0b}) (see also \cite{Feynman}). The propagator
$K(x,x^{\prime};t)=<x|\exp(-iHt/\hbar)|x^{\prime}>$ is the probability
amplitude for particle transition from the initial spacetime point
($x^{\prime},0$) to the final point ($x,t$) by means of all possible paths. It
provides full information on the particle's dynamics and resolves the
corresponding time-dependent Schr\"{o}dinger equation (\ref{0}). According to
\cite{Los 2010}, the time-dependent retarded propagator $K(t)=\theta
(t-t^{\prime})\exp(-\frac{i}{\hbar}Ht)$ can be represented as%
\begin{equation}
K(t)=\theta(t)\frac{i}{2\pi}%
{\displaystyle\int\limits_{-\infty}^{\infty}}
dEe^{-\frac{i}{\hbar}Et}(\frac{1}{E-H+i\varepsilon}-\frac{1}{E-H-i\varepsilon
}),\varepsilon\rightarrow+0,\label{3b}%
\end{equation}
where $H$ is the time-independent Hamiltonian of the system under
consideration. Equation (\ref{3b}) follows either from the contour integration
in the complex plane or from the identity%
\begin{equation}
\frac{1}{E-H\pm i\varepsilon}=P\frac{1}{E-H}\mp i\pi\delta(E-H),\label{3c}%
\end{equation}
where $P$ is the symbol of the integral principal value. In the space
representation (\ref{3b}) reads%
\begin{equation}
K(x,x^{\prime};t)=\theta(t)%
{\displaystyle\int\limits_{-\infty}^{\infty}}
e^{-\frac{i}{\hbar}Et}A(x,x^{\prime};E)dE.\label{4}%
\end{equation}
Here $A(x,x^{\prime};E)$ is the spectral density matrix
\begin{align}
A(x,x^{\prime};E) &  =\frac{i}{2\pi}\left[  G^{+}(x,x^{\prime};E)-G^{-}%
(x,x^{\prime};E)\right]  ,\nonumber\\
G^{+}(x,x^{\prime};E) &  =<x|\frac{1}{E-H+i\varepsilon}|x^{\prime}%
>,G^{-}(x,x^{\prime};E)=\left[  G^{+}(x^{\prime},x;E)\right]  ^{\ast
},\varepsilon\rightarrow+0,\label{5}%
\end{align}
determined by the matrix elements of the retarded $G^{+}(E)$ and advanced
$G^{-}(E)$ energy-dependent operator Green functions $G^{\pm}=(E-H\pm
i\varepsilon)^{-1}$, which are analytical in the upper and lower half-planes
of the complex energy $E$, respectively. The propagator in the form of
(\ref{4}) is a useful tool for calculations within the multiple-scattering
theory (MST) perturbation expansion if the Hamiltonian can be split as
$H=H_{0}+H_{i}$, where $H_{0}$ describes a free motion and $H_{i}$ is the
scattering potential. Note, that in this case one would not need rely on the
standard (often cumbersome) matching procedure characteristic of the picture
when a wave (representing a particle) is reflected from and transmitted
through the potential (\ref{1}). On the other hand, the introduction of the
scattering potential $H_{i}$ corresponds to the natural picture of the
particle scattering at the potential jumps at $x=0 $ and $x=d$.

We showed in \cite{Los 2010} that the Hamiltonian corresponding to the
energy-conserving processes of scattering at potential steps can be presented
as
\begin{align}
H &  =H_{0}+H_{i}(x;E),\nonumber\\
H_{i}(x;E) &  =%
{\displaystyle\sum\limits_{s}}
H_{i}^{s}(E)\delta(x-x_{s}).\label{7}%
\end{align}
Here, $H_{i}(x;E)$ describes the perturbation of the "free" particle motion
(defined by $H_{0}=-\frac{\hbar^{2}}{2m}\frac{\partial^{2}}{\partial x^{2}} $)
localized at the potential steps with coordinates $x_{s}$ (in the case of the
potential (\ref{1}), there are two potential steps at $x_{s}=0$ and $x_{s}=d$)%
\begin{align}
H_{i>}^{s}(E) &  =\frac{i\hbar}{2}[v_{>}^{s}(E)-v_{<}^{s}(E)],\nonumber\\
H_{i<}(E) &  =\frac{i\hbar}{2}[v_{<}^{s}(E)-v_{>}^{s}(E)],\nonumber\\
H_{i><}(E) &  =\frac{2i\hbar v_{>}^{s}(E)v_{<}^{s}(E)}{[\sqrt{v_{>}^{s}%
(E)}+\sqrt{v_{<}^{s}(E)}]^{2}},\label{8}%
\end{align}
where $H_{i>(<)}^{s}(E)$ is the reflection (from the potential step at
$x=x_{s}$) potential amplitude, the index $>(<)$ indicates the side on which
the particle approaches the interface at $x=x_{s}$: right ($>$) or left ($<$);
$H_{i><}^{s}(E)$ is the transmission potential amplitude, and the velocities
$v_{>(<)}^{s}(E)=\hbar k_{>(<)}^{s}(E)/m,s\in\{0,d\}$ ($k_{>(<)}^{s}(E)$ are
given by (\ref{2})).

The perturbation expansion for the retarded Green function $G^{+}(x,x^{\prime
};E)$ in the case of the rectangular potential (\ref{1}), which can be
effectively represented by the two-step effective scattering Hamiltonian
(\ref{7}), reads for different source (given by $x^{\prime}$) and destination
(determined by $x$) areas of interest as%
\begin{align}
G^{+}(x,x^{\prime};E) &  =G_{0}^{+}(x,d;E)T^{+}(E)G_{0}^{+}(0,x^{\prime
};E),x^{\prime}<0,x>d,\nonumber\\
G^{+}(x,x^{\prime};E) &  =G_{0}^{+}(x,0;E)T^{+}(E)G_{0}^{+}(d,x^{\prime
};E),x^{\prime}>d,x<0,\nonumber\\
G^{+}(x,x^{\prime};E) &  =G_{0}^{+}(x,0;E)T^{\prime+}(E)G_{0}^{+}(0,x^{\prime
};E)+G_{0}^{+}(x,d;E)R^{\prime+}(E)G_{0}^{+}(0,x^{\prime};E),x^{/}%
<0,0<x<d,\nonumber\\
G^{+}(x,x^{\prime};E) &  =G_{0}^{+}(x,0;E)T^{\prime+}(E)G_{0}^{+}(0,x^{\prime
};E)+G_{0}^{+}(x,0;E)R^{\prime+}(E)G_{0}^{+}(d,x^{\prime};E),0<x^{/}%
<d,x<0,\nonumber\\
G^{+}(x,x^{\prime};E) &  =G_{0}^{+}(x,x^{\prime};E)+G_{0}^{+}(x,0;E)R^{+}%
(E)G_{0}^{+}(0,x^{\prime};E),x^{\prime}<0,x<0,\label{9}%
\end{align}
where the transmission and reflection matrices are
\begin{align}
T^{+}(E) &  =\frac{T_{><}^{d+}(E)G_{0}^{+}(d,0;E)T_{><}^{0+}(E)}{D^{+}%
(E)},\nonumber\\
T^{\prime+}(E) &  =\frac{T_{><}^{0+}(E)}{D^{+}(E)},R^{\prime+}(E)=T_{<}%
^{d+}(E)G_{0}^{+}(d,0;E)T^{\prime+}(E),\nonumber\\
R^{+}(E) &  =T_{<}^{0+}(E)+\frac{T_{><}^{0+}(E)G_{0}^{+}(0,d;E)T_{<}%
^{d+}(E)G_{0}^{+}(d,0;E)T_{><}^{0+}(E)}{D^{+}(E)},\nonumber\\
D^{+}(E) &  =1-T_{<}^{d+}(E)G_{0}^{+}(d,0;E)T_{>}^{0+}(E)G_{0}^{+}%
(0,d;E).\label{10}%
\end{align}
The one-dimensional retarded Green function $G_{0}^{+}(x,x^{\prime};E)$
corresponding to a free particle moving in constant potential $V(x)=0$ or
$V(x)=U($or $\Delta)$ is (see, e.g. \cite{Economou})%
\begin{align}
G_{0}^{+}(x,x^{\prime};E) &  =\frac{m}{i\hbar^{2}k(E)}\exp[ik(E)|x-x^{\prime
}|],V(x)=0,\nonumber\\
G_{0}^{+}(x,x^{\prime};E) &  =\frac{m}{i\hbar^{2}k_{u(\Delta)}(E)}%
\exp[ik_{u(\Delta)}(E)|x-x^{\prime}|],V(x)=U(\text{or }\Delta),\label{11}%
\end{align}
where the wave numbers are determined by (\ref{2}). The scattering (at the
step located at $x=x_{s}$) t-matrices are defined by the following
perturbation expansion%
\begin{align}
T^{s}(E) &  =H_{i}^{s}(E)+H_{i}^{s}(E)G_{0}(x_{s},x_{s};E)H_{i}^{s}%
(E)+\ldots\nonumber\\
&  =\frac{H_{i}^{s}(E)}{1-G_{0}(x_{s},x_{s};E)H_{i}^{s}(E)},\label{12}%
\end{align}
where $H_{i}^{s}(E)$ and the interface Green function $G_{0}(x_{s},x_{s};E)$
are defined differently for reflection and transmission processes \cite{Los
2010}: the step-localized effective potential is given by Eq. (\ref{8}) and
the retarded Green functions at the interface for the considered reflection
and transmission processes are, correspondingly,
\begin{align}
G_{0>(<)}^{+}(x_{s},x_{s};E) &  =1/i\hbar v_{>(<)}^{s}(E)\nonumber\\
G_{0><}^{+}(x_{s},x_{s};E) &  =1/i\hbar\sqrt{v_{>}^{s}(E)v_{<}^{s}%
(E)}\label{13}%
\end{align}
in accordance with (\ref{11}).

From (\ref{8}), (\ref{12}) and (\ref{13}), we have for the reflection
$T_{>(<)}^{s+}(E)$ and transmission $T_{><}^{s+}(E)$ t-matrices, used in
(\ref{10}) ($s\in\{0,d\}$) and corresponding to the retarded Green function
and scattering at the interface located at $x=x_{s}\in\{0,d\}$:
\begin{align}
T_{>(<)}^{s+}(E) &  =i\hbar v_{>(<)}^{s}r_{>(<)}^{s},\nonumber\\
T_{><}^{s+}(E) &  =i\hbar\sqrt{v_{>}^{s}v_{<}^{s}}t^{s},\label{14}%
\end{align}
where $r_{>(<)}^{s}(E)$ and $t^{s}(E)$ are the standard amplitudes for
reflection to the right (left) of the potential step at $x=x_{s}$ and
transmission through this step%
\begin{align}
r_{>}^{s}(E) &  =\frac{k_{>}^{s}-k_{<}^{s}}{k_{>}^{s}+k_{<}^{s}},r_{<}%
^{s}(E)=\frac{k_{<}^{s}-k_{>}^{s}}{k_{>}^{s}+k_{<}^{s}},\nonumber\\
t^{s}(E) &  =\frac{2\sqrt{k_{>}^{s}k_{<}^{s}}}{k_{>}^{s}+k_{<}^{s}},\label{15}%
\end{align}
and the argument $E$ is omitted for brevity.

Now, using (\ref{2}), (\ref{9}), (\ref{10}), (\ref{11}), (\ref{14}) and
(\ref{15}), we can obtain the Green function $G^{+}(x,x^{\prime};E)$ for the
spatial domains considered in (\ref{9}) (see \cite{Los 2013})%

\begin{align}
G^{+}(x,x^{\prime};E) &  =\frac{m}{i\hbar^{2}\sqrt{kk_{\Delta}}}e^{ik_{\Delta
}(x-d)}t(E)e^{-ikx^{\prime}},x^{\prime}<0,x>d,\nonumber\\
G^{+}(x,x^{\prime};E) &  =\frac{m}{i\hbar^{2}\sqrt{kk_{\Delta}}}%
e^{-ikx}t(E)e^{ik_{\Delta}(x^{\prime}-d)},x^{\prime}>d,x<0,\nonumber\\
G^{+}(x,x^{\prime};E) &  =\frac{m}{i\hbar^{2}\sqrt{kk_{u}}}\left[  e^{ik_{u}%
x}t^{\prime}(E)e^{-ikx^{\prime}}+e^{-ik_{u}x}r^{\prime}(E)e^{-ikx^{\prime}%
}\right]  ,x^{\prime}<0,0<x<d,\nonumber\\
G^{+}(x,x^{\prime};E) &  =\frac{m}{i\hbar^{2}\sqrt{kk_{u}}}\left[
e^{-ikx}t^{\prime}(E)e^{ik_{u}x^{\prime}}+e^{-ikx}r^{\prime}(E)e^{-ik_{u}%
x^{\prime}}\right]  ,x<0,0<x^{\prime}<d,\nonumber\\
G^{+}(x,x^{\prime};E) &  =\frac{m}{i\hbar^{2}k}\left[  e^{ik\left\vert
x-x^{\prime}\right\vert }+r(E)e^{-ik(x+x^{\prime})}\right]  ,x<0,x^{\prime
}<0,\label{16}%
\end{align}
where the transmission and reflection amplitudes are defined as%
\begin{align}
t(E) &  =\frac{4\sqrt{kk_{\Delta}}k_{u}e^{ik_{u}d}}{d(E)},t^{\prime}%
(E)=\frac{2\sqrt{kk_{u}}(k_{\Delta}+k_{u})}{d(E)},\nonumber\\
r^{\prime}(E) &  =\frac{2\sqrt{kk_{u}}(k_{u}-k_{\Delta})e^{2ik_{u}d}}%
{d(E)},r(E)=\frac{(k-k_{u})(k_{\Delta}+k_{u})-(k+k_{u})(k_{\Delta}%
-k_{u})e^{2ik_{u}d}}{d(E)},\nonumber\\
d(E) &  =(k+k_{u})(k_{\Delta}+k_{u})-(k-k_{u})(k_{\Delta}-k_{u})e^{2ik_{u}%
d}.\label{17}%
\end{align}
Using the same approach, it is not difficult to obtain the Green function
$G^{+}(x,x^{\prime};E)$ for other areas of arguments $x$ and $x^{\prime}$.

In accordance with the obtained results for Green's functions, we consider the
situation when a particle, given originally by a wave packet localized to the
left of the potential area, i.e. at $x^{\prime}<0$, moves towards the
potential (\ref{1}). We also choose $\Delta\geqslant0$, which corresponds to
the case when, e.g., the spin-up electrons of the left magnetic layer
($x^{\prime}<0$) move through the nonmagnetic spacer to the right magnetic
layer ($x>d$) aligned either in parallel ($\Delta=0$) or antiparallel
($\Delta>0$) to the left magnetic layer. At the same time, the amplitude $U$
in the potential (\ref{1}) may acquire both positive (barrier) and negative
(well) values.

From Eqs. (\ref{16}) we see that $G^{+}(x,x^{\prime};E)=G^{+}(x^{\prime}%
,x;E)$, and, therefore, the advanced Green function $G^{-}(x,x^{\prime
};E)=\left[  G^{+}(x^{\prime},x;E)\right]  ^{\ast}=\left[  G^{+}(x,x^{\prime
};E)\right]  ^{\ast}$ (see, e.g. \cite{Economou}). Thus, the transmission
amplitude (\ref{4}) is determined by the imaginary part of the Green function
and can be written as%
\begin{equation}
K(x,x^{\prime};t)=-\theta(t)\frac{1}{\pi}%
{\displaystyle\int\limits_{-\infty}^{\infty}}
dEe^{-\frac{i}{\hbar}Et}\operatorname{Im}G^{+}(x,x^{\prime};E).\label{18}%
\end{equation}

Formulas (\ref{16}) - (\ref{18}) present the exact solution for the particle
propagator in the presence of the potential (\ref{1}). It should be kept in
mind that the wave numbers (\ref{2}) and, therefore, the quantities $t(E)$,
$t^{\prime}(E)$, $r^{\prime}(E)$, and $r(E)$ in (\ref{17}) are different in
the $%
{\displaystyle\int\limits_{-\infty}^{0}}
dE$ and $%
{\displaystyle\int\limits_{0}^{\infty}}
dE$ energy integration areas: in the former case, $k(E)$ and $k_{\Delta}(E)$
($\Delta\geqslant0$) should be replaced with $i\overline{k}(E)$ and
$i\overline{k}_{\Delta}(E)$, where $\overline{k}(E)=\sqrt{-2mE/\hbar^{2}}$
($E<0$) and $\overline{k}_{\Delta}(E;\mathbf{k}_{||})=\sqrt{2m(\Delta
-E)/\hbar^{2}}$. At the same time, for energies $E<0$, the wave number
$k_{u}=i\overline{k}_{u}$, $\overline{k}_{u}=\sqrt{2m(U-E)/h^{2}}$, for $U>0$
(barrier), but for $U<0$ it is real, i.e. $k_{u}=\sqrt{2m(E+\left\vert
U\right\vert )/\hbar^{2}}$, if $E>-\left\vert U\right\vert $ and
$k_{u}=i\overline{k}_{u}$, $\overline{k}_{u}=\sqrt{-2m(E+\left\vert
U\right\vert )/\hbar^{2}}$ if $E<-\left\vert U\right\vert $. It follows that
the "free" Green function $G_{0}^{+}(x,x^{\prime};E)=\frac{m}{i\hbar^{2}%
k}e^{ik\left\vert x-x^{\prime}\right\vert }$ is real in the energy interval
($-\infty\div0$) and, therefore, does not contribute in this interval to the
corresponding "free" propagator $K_{0}(x,x^{\prime};t)$ defined by (\ref{18}).
It is also remarkable that for energies $E<0$ the imaginary parts of the Green
functions vanish in all spatial regions, as is seen from definitions
(\ref{16}) and (\ref{17}) (e.g., $\operatorname{Im}t(E)=0$ and
$\operatorname{Im}r(E)=0$ for $E<0$). Therefore, the energy interval
($-\infty\div0$) does not contribute to the propagation of the particles
through the potential well/barrier region. Thus, we have for $t>0$%

\begin{align}
K(x,x^{\prime};t)  & =\frac{1}{\pi\hbar}%
{\displaystyle\int\limits_{0}^{\infty}}
\frac{dEe^{-\frac{i}{\hbar}Et}}{\sqrt{v(E)}}\operatorname{Re}\left[
\frac{t(E)e^{ik_{\Delta}(E)(x-d)}e^{-ik(E)x^{\prime}}}{\sqrt{v_{\Delta}(E)}%
}\right]  ,x^{\prime}<0,x>d,\nonumber\\
K(x,x^{\prime};t)  & =\frac{1}{\pi\hbar}%
{\displaystyle\int\limits_{0}^{\infty}}
\frac{dEe^{-\frac{i}{\hbar}Et}}{\sqrt{v(E)}}\operatorname{Re}\left\{
\frac{e^{-ik(E)x^{\prime}}\left[  t^{\prime}(E)e^{ik_{u}(E)x}+r^{\prime
}(E)e^{-ik_{u}(E)x}\right]  }{\sqrt{v_{u}(E)}}\right\}  ,x^{\prime
}<0,0<x<d,\nonumber\\
K(x,x^{\prime};t)  & =\frac{1}{\pi\hbar}%
{\displaystyle\int\limits_{0}^{\infty}}
\frac{dEe^{-\frac{i}{\hbar}Et}}{v(E)}\operatorname{Re}[e^{ik(E)\left\vert
x-x^{\prime}\right\vert }+r(E)e^{-ik(E)(x+x^{\prime})}],x^{\prime
}<0,x<0,\label{19}%
\end{align}
where the velocities $v(E)$, $v_{u}(E)$ and $v_{\Delta}(E)$ are defined by
(\ref{2}) with the multiplier $\hbar/m$.

It is easy to verify that the integration over $E$ of the first term in the
last line of (\ref{19}) results in the known formula for the space-time
propagator for a freely moving particle
\begin{equation}
K_{0}(x,x^{\prime};t)=(\frac{m}{2\pi i\hbar t})^{1/2}\exp\left[
\frac{im(x-x^{\prime})^{2}}{2\hbar t}\right]  ,x<0,x^{\prime}<0.\label{20}%
\end{equation}

The obtained results (\ref{19}) for the particle propagator completely resolve
(by means of Eq. (\ref{0b})) the time-dependent Schr\"{o}dinger equation for a
particle moving under the influence of the rectangular potential (\ref{1}).
The form of this solution (integrals from the elementary functions) is
convenient for numerical visualization. Further application of these results
to the calculation of the TOA and dwell time as well as of the probability
density of finding a particle in different spatial areas as a function of time
with account of the forward- and backward-moving components of the wave
function and their interference can be found in our earlier papers \cite{Los
2010,Los 2011,Los 2012, Los 2013}.

\section{Application to the density matrix}

The equilibrium non-normalized density operator (propagator in the temperature
domain) $\rho(\beta)=\exp(-\beta H)$ can likewise be expressed in terms of the
resolvent operator $(E-H)^{-1}$ (see (\ref{3b}))%
\begin{equation}
\rho(\beta)=\exp(-\beta H)=\frac{i}{2\pi}\int\limits_{-\infty}^{\infty
}dEe^{-\beta E}(\frac{1}{E-H+i\varepsilon}-\frac{1}{E-H-i\varepsilon
}).\label{21}%
\end{equation}
Particularly, in the coordinate representation we have for the density matrix
(see (\ref{0c}))%
\begin{equation}
\rho(x,x^{\prime};\beta)=\int\limits_{-\infty}^{\infty}e^{-\beta
E}A(x,x^{\prime};E)dE,\label{22}%
\end{equation}
where $A(x,x^{\prime};E)$ is given by (\ref{5}). Thus, the density matrix
$\rho(x,x^{\prime};\beta)$ follows from the propagator (\ref{4}) by the
substitution $t\rightarrow-i\hbar\beta$ ($\beta=1/k_{B}T$). From the
properties (\ref{5}) we see that the density matrix (\ref{22}) is
self-adjoint. The density operator (\ref{21}) satisfies the Bloch equation
(\ref{0d}).

Thus, shifting to the imaginary "time" ($t\rightarrow-i\hbar\beta$), we obtain
the exact density matrix $\rho(x,x^{\prime};\beta)$ in the various considered
(relative to the potential (\ref{1}) area) spatial regions, i.e.
\begin{equation}
\rho(x,x^{\prime};\beta)=K(x,x^{\prime};-i\hbar\beta),\label{23}%
\end{equation}
where $K(x,x^{\prime};t=-i\hbar\beta)$ is given by (\ref{19}). In particular,
we obtain from (\ref{20}) the known result for the "free" density matrix
\begin{equation}
\rho_{0}(x,x^{\prime};\beta)=K_{0}(x,x^{\prime};-i\hbar\beta)=(\frac{m}%
{2\pi\hbar^{2}\beta})^{1/2}\exp[-\frac{m(x-x^{\prime})^{2}}{2\hbar^{2}\beta
}].\label{24}%
\end{equation}

Using the same approach, it is not difficult to obtain the propagator
$\rho(x,x^{\prime};\beta)$ for other (than in (\ref{19})) areas of the
arguments $x$ and $x^{\prime}$. Again, it is important to note that the
negative energy half line ($-\infty\div0$), corresponding to the evanescent
states does not contribute to the propagator (\ref{22}). The diagonal element
$K(x,x;\beta)$ ($x=x^{\prime}$ can be put only in the last line of (\ref{19}))
defines the density of particles per unit length at the point $x<0$ to the
left of the potential (\ref{1}). The nondiagonal elements $K(x,x^{\prime
};\beta)$ of (\ref{19}) are related to the quantum mechanical interference
effects. Particularly, they are responsible for particle tunneling through the
barrier and also can be attributed to the phase correlation of the states
$|x>$ and $|x^{\prime}>$.

Equations (\ref{19}) and (\ref{23}) provide an exact solution for the particle
density matrix in the presence of the rectangular potential (\ref{1}) in terms
of integrals of elementary functions. It is convenient (e.g., for numerical
visualization of the obtained results) to shift to dimensionless variables. As
seen from (\ref{1}), (\ref{2}) and (\ref{19}), there are the natural spatial
scale $d$ and the energy scale $E_{d}=\hbar^{2}/2md^{2}$ (the energy
uncertainty due to particle localization within a potential range of width
$d$). Then, the density matrix (\ref{23}) in the different spatial regions can
be presented in the dimensionless variables as%
\begin{align}
\rho(\widetilde{x},\widetilde{x}^{\prime};\widetilde{\beta})  & =\frac{1}{2\pi
d}\int\limits_{0}^{\infty}\frac{d\widetilde{E}e^{-\widetilde{\beta}%
\widetilde{E}}}{\widetilde{E}^{1/4}}\operatorname{Re}\left[  \frac
{\widetilde{t}(\widetilde{E})e^{i\sqrt{\widetilde{E}-\widetilde{\Delta}%
}(\widetilde{x}-1)}e^{-i\sqrt{\widetilde{E}}\widetilde{x}^{\prime}}%
}{(\widetilde{E}-\widetilde{\Delta})^{1/4}}\right]  ,\widetilde{x}^{\prime
}<0,\widetilde{x}>1,\nonumber\\
\rho(\widetilde{x},\widetilde{x}^{\prime};\widetilde{\beta})  & =\frac{1}{2\pi
d}\int\limits_{0}^{\infty}\frac{d\widetilde{E}e^{-\widetilde{\beta}%
\widetilde{E}}}{\widetilde{E}^{1/4}}\operatorname{Re}\left\{  \frac{\left[
\widetilde{t}^{\prime}(\widetilde{E})e^{i\sqrt{\widetilde{E}-\widetilde{U}%
}\widetilde{x}}+\widetilde{r}^{\prime}(\widetilde{E})e^{-i\sqrt{\widetilde
{E}-\widetilde{U}}\widetilde{x}}\right]  e^{-i\sqrt{\widetilde{E}}%
\widetilde{x}^{\prime}}}{(\widetilde{E}-\widetilde{U})^{1/4}}\right\}
,\widetilde{x}^{\prime}<0,0<\widetilde{x}<1,\nonumber\\
\rho(\widetilde{x},\widetilde{x}^{\prime};\widetilde{\beta})  & =\frac
{1}{2\sqrt{\pi\widetilde{\beta}}d}\exp[-(\widetilde{x}-\widetilde{x}^{\prime
})^{2}/4\widetilde{\beta}]+\frac{1}{2\pi d}\int\limits_{0}^{\infty}%
\frac{d\widetilde{E}e^{-\widetilde{\beta}\widetilde{E}}}{\sqrt{\widetilde{E}}%
}\operatorname{Re}[\widetilde{r}(\widetilde{E})e^{=i\sqrt{\widetilde{E}%
}(\widetilde{x}+\widetilde{x}^{\prime})}],\widetilde{x}^{\prime}%
<0,\widetilde{x}<0,\label{25}%
\end{align}
where%
\begin{align}
\widetilde{t}(\widetilde{E})  & =\frac{4\widetilde{E}^{1/4}(\widetilde
{E}-\widetilde{\Delta})^{1/4}\sqrt{\widetilde{E}-\widetilde{U}}e^{i\sqrt
{\widetilde{E}-\widetilde{U}}}}{\widetilde{d}(\widetilde{E})},\nonumber\\
\widetilde{t}^{\prime}(\widetilde{E})  & =\frac{2\widetilde{E}^{1/4}%
(\widetilde{E}-\widetilde{U})^{1/4}(\sqrt{\widetilde{E}-\widetilde{\Delta}%
}+\sqrt{\widetilde{E}-\widetilde{U}})}{\widetilde{d}(\widetilde{E}%
)},\nonumber\\
\widetilde{r}^{\prime}(\widetilde{E})  & =\frac{2\widetilde{E}^{1/4}%
(\widetilde{E}-\widetilde{U})^{1/4}(\sqrt{\widetilde{E}-\widetilde{U}}%
-\sqrt{\widetilde{E}-\widetilde{\Delta}})e^{2i\sqrt{\widetilde{E}%
-\widetilde{U}}}}{\widetilde{d}(\widetilde{E})},\nonumber\\
\widetilde{r}(\widetilde{E})  & =\frac{(\sqrt{\widetilde{E}}-\sqrt
{\widetilde{E}-\widetilde{U}})(\sqrt{\widetilde{E}-\widetilde{\Delta}}%
+\sqrt{\widetilde{E}-\widetilde{U}})-(\sqrt{\widetilde{E}}+\sqrt{\widetilde
{E}-\widetilde{U}})(\sqrt{\widetilde{E}-\widetilde{\Delta}}-\sqrt
{\widetilde{E}-\widetilde{U}})e^{2i\sqrt{\widetilde{E}-\widetilde{U}}}%
}{\widetilde{d}(\widetilde{E})},\nonumber\\
\widetilde{d}(\widetilde{E})  & =(\sqrt{\widetilde{E}}+\sqrt{\widetilde
{E}-\widetilde{U}})(\sqrt{\widetilde{E}-\widetilde{\Delta}}+\sqrt
{\widetilde{E}-\widetilde{U}})-(\sqrt{\widetilde{E}}-\sqrt{\widetilde
{E}-\widetilde{U}})(\sqrt{\widetilde{E}-\widetilde{\Delta}}-\sqrt
{\widetilde{E}-\widetilde{U}})e^{2i\sqrt{\widetilde{E}-\widetilde{U}}%
},\label{26}%
\end{align}
and $\widetilde{E}=E/E_{d}$, $\widetilde{U}=U/E_{d}$, $\widetilde{\Delta
}=\Delta/E_{d}$, $\widetilde{\beta}=E_{d}/k_{B}T=\widetilde{T}/T$,
$\widetilde{T}=E_{d}/k_{B}$, $\widetilde{x}=x/d$, $\widetilde{x}^{\prime
}=x^{\prime}/d$.

We will visualize the results given by Eqs. (\ref{25}) and (\ref{26}) for
several specific values of the relevant parameters. For an electron and the
potential width $d=10^{-7}cm$ ($1nm$), the characteristic energy $E_{d}%
\sim3\cdot10^{-2}ev$ and the characteristic temperature $\widetilde{T}%
=E_{d}/k_{B}\backsim3\cdot10^{2}K$.

We will perform the numerical modeling of the density matrix (\ref{25}) with
the symmetric rectangular potential (\ref{1}) when $\Delta=0$ (in this case
the transition and reflection amplitudes (\ref{26}) simplify essentially). To
secure a rapid convergence of the integrals in (\ref{25}), we consider low
enough temperatures, i.e. put $\widetilde{\beta}=10$ ($k_{B}T\ll E_{d}$).
Figure 1 shows the diagonal element of the density matrix $\rho(\widetilde
{x},\widetilde{x};\widetilde{\beta})$ (the last line in (\ref{25})) at
$\widetilde{x}=-2$, i.e. the probability density to find a particle at this
spatial point to the left of the barrier as a function of the potential well
modulus $\left\vert \widetilde{U}\right\vert $ ($\widetilde{U}=0\div-300$). We
see that in this case the density matrix $\rho(\widetilde{x},\widetilde
{x};\widetilde{\beta})$ exhibits a series of maximums and minimums. This can
be explained by the formation of the resonance levels above the well if the
condition $\widetilde{E}+\left\vert \widetilde{U}\right\vert =\pi^{2}n^{2}$
($n$ is integer, $n=1,2,\ldots$) holds. With such a condition we have the
reflection amplitude $\widetilde{r}(\widetilde{E})=0$ and the transmission
amplitude $\widetilde{t}(\widetilde{E})=\pm1$. As at low temperatures
($\widetilde{\beta}=10$) the main contribution to the integral over
$\widetilde{E}$ comes from the small (close to zero) energies, the positions
of jumps at Fig. 1 approximately follow the relation $\left\vert \widetilde
{U}\right\vert =\pi^{2}n^{2}$ ($n=1,2,\ldots$).%

\begin{figure}
[ptb]
\begin{center}
\includegraphics[
height=2.2554in,
width=3.659in
]%
{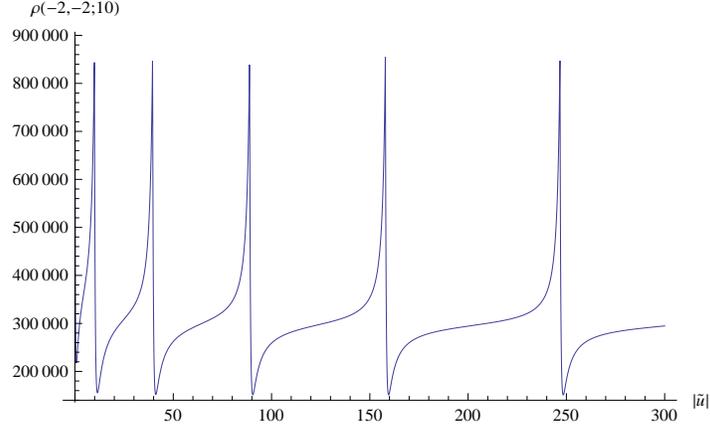}%
\caption{The diagonal element $\rho(-2,-2;10)$ as a function of the depth of
the potential well $\left\vert \widetilde{U}\right\vert $.}%
\label{Fig1}%
\end{center}
\end{figure}

The same diagonal element $\rho(-2,-2;10)$ as a function of the height of the
potential barrier $\widetilde{U}=0\div300$ behaves quite different from the
case of the potential well and is shown in Fig. 2. One can see that the
particle probability density at the given point to the left of the barrier
$\widetilde{x}=-2$ exhibits at first a steep fall with the potential barrier
growth and then it changes slowly with $\widetilde{U}$.
\begin{figure}
[ptb]
\begin{center}
\includegraphics[
height=2.1715in,
width=3.659in
]%
{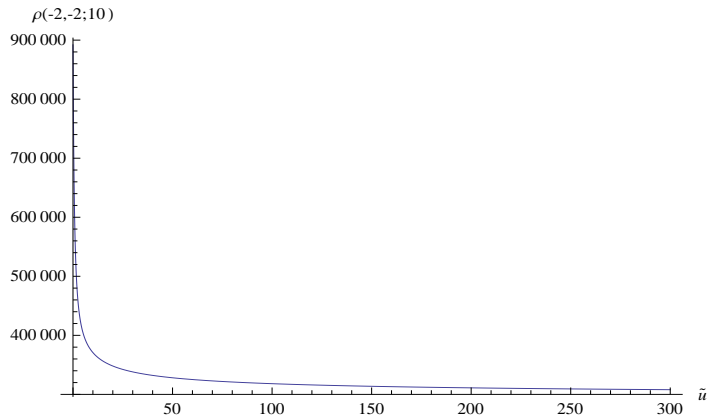}%
\caption{The same (as in Fig. 1) diagonal element of the density matrix as a
function of the potential barrier height $\widetilde{U}$.}%
\label{Fig2}%
\end{center}
\end{figure}

We will evaluate the nondiagonal elements of the density matrix $\rho
(\widetilde{x},\widetilde{x}^{\prime};\widetilde{\beta})$ for points at
different sides of the well/barrier (the first line in (\ref{25})) as a
function of the potential parameter $\widetilde{U}$. Thus, we put
$\widetilde{x}^{\prime}=-10$ (before the potential), $\widetilde{x}=2$ (beyond
the potential) and $\widetilde{\beta}=10$ (as for Figs. (\ref{1}) and
(\ref{2})). Figure 3 exhibits the picks from the density matrix for the case
of the potential well with $\widetilde{U}=0\div-300$ at the resonance values
of $\left\vert \widetilde{U}\right\vert $ which correspond to the minima in
Fig. 1.%

\begin{figure}
[ptb]
\begin{center}
\includegraphics[
height=2.0738in,
width=3.659in
]%
{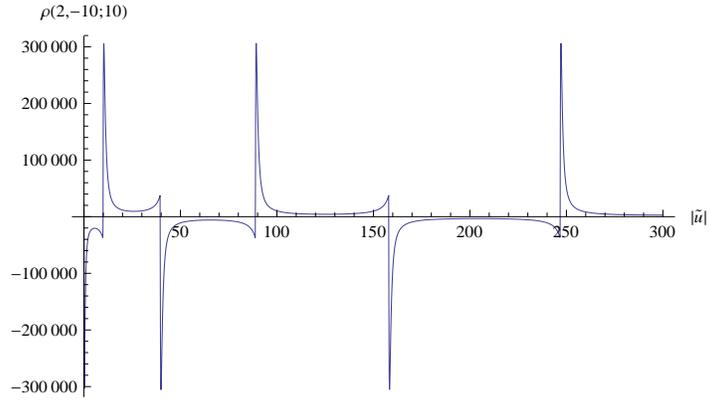}%
\caption{The nondiagonal element of the density matrix $\rho(2,-10;10)$ as a
function of the potential well depth $\left\vert \widetilde{U}\right\vert $.}%
\label{Fig3}%
\end{center}
\end{figure}
We see that in this case ($\widetilde{U}<0$) the density matrix nondiagonal
elements can acquire both positive and negative values. Note that at
$\widetilde{U}=0$ the density matrix reduces to the free density matrix
(\ref{24}) and therefore is positive (the seemingly negative value of the
density matrix close to $\widetilde{U}=0$ in Fig. 3 is due to small resolution
on the $\left\vert \widetilde{U}\right\vert $-axis; calculation on the smaller
scale near the point $\widetilde{U}=0$ shows that at $\widetilde{U}=0$ the
density matrix is positive). Thus, Fig. 3 demonstrates the jumps of the
quantum coherence between the particle states before ($x^{\prime}<0$) and
beyond ($x>0$) the potential well at the resonance particle transmission
through the potential well. At the values of $\left\vert \widetilde
{U}\right\vert $ that do not satisfy the resonance condition this quantum
coherence is small.

The nondiagonal matrix element $\rho(2,-10;10)$ as a function of the potential
barrier height ($\widetilde{U}=0\div100$) is shown in Fig. 4. We see that the
quantum coherence between the states on the different sides of the barrier
goes quickly enough to zero along with the barrier height.%

\begin{figure}
[ptb]
\begin{center}
\includegraphics[
height=2.2278in,
width=3.659in
]%
{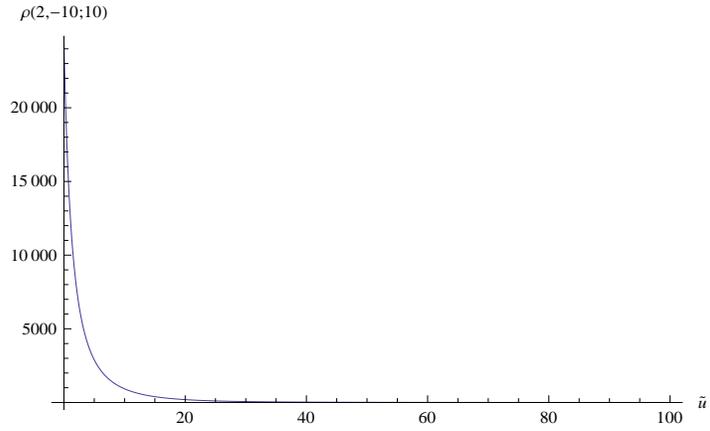}%
\caption{The dependence of $\rho(2,-10;10)$ on the potential barrier height
$\widetilde{U}$.}%
\label{Fig4}%
\end{center}
\end{figure}

\section{Diffusion-like equation}

As mentioned in the Introduction, the time-dependent Schr\"{o}dinger equation
(\ref{0}) becomes equivalent to the parabolic diffusion-like equation
(\ref{0e}) if one makes the substitutions $t\rightarrow-it$ and $\hbar
\rightarrow2mD$, where $D$ is a (diffusion) constant. Thus, we can immediately
obtain from Eqs. (\ref{19}) a solution to the diffusion equation (\ref{0e})
for the initial condition $Q(x;0)=\delta(x-x^{\prime})$. In the dimensionless
variables, this solution (a propagator) is given by Eqs. (\ref{25}),
(\ref{26}) with the following substitutions
\begin{align}
\rho(\widetilde{x},\widetilde{x}^{\prime};\widetilde{\beta})  & \rightarrow
Q(\widetilde{x},\widetilde{x}^{\prime};\overline{t}),\widetilde{\beta
}\rightarrow\overline{t},\widetilde{E}\rightarrow\overline{E},\widetilde
{U}\rightarrow\overline{U},\widetilde{\Delta}\rightarrow\overline{\Delta
},\nonumber\\
\overline{t}  & =t/t_{D},\overline{E}=E/E_{D},\overline{U}=U/E_{D}%
,\overline{\Delta}=\Delta/E_{D},\nonumber\\
t_{D}  & =d^{2}/D,E_{D}=2mD^{2}/d^{2},\label{27}%
\end{align}
where $t_{D}$ and $E_{D}$ are obtained from $t_{d}$ and $E_{d}$ of the
previous section by the substitution $\hbar\rightarrow2mD$. The introduced
characteristic time $t_{D}$, as it follows from the definition (\ref{27}), can
be interpreted as the time needed for a particle to diffuse over the distance
$d$ (a potential (\ref{1}) width) with the diffusion coefficient $D$. The
characteristic energy $E_{D}=2mD^{2}/d^{2}=2md^{2}/t_{D}^{2}=4\frac{mv_{D}%
^{2}}{2}$ is proportional to the kinetic energy of a particle moving with the
average velocity $v_{D}=d/t_{D}$. Therefore, as in the previous section, we
can numerically model the solution $Q(\widetilde{x},\widetilde{x}^{\prime
};\overline{t})$ defined by Eqs. (\ref{25}), (\ref{26}) (with the
substitutions (\ref{27})) in the different spatial points $\widetilde{x}=x/d$
and $\widetilde{x}^{\prime}=x/d$.

Note that at $\overline{V}(x)>0$ the solution to the diffusion equation
(\ref{0e}) is positive, $Q(\widetilde{x},\widetilde{x}^{\prime};\overline
{t})\geq0$ (see \cite{Kac}) and can be viewed as the density of particles in
the point $\widetilde{x}$ at the moment of time $\overline{t}$ when the
"diffusion with the holes" starts at the point $\widetilde{x}^{\prime}$. The
latter term was introduced by Kac because in the points, where the potential
(\ref{1}) $\overline{V}(x)\neq0$ ($\overline{V}(x)>0$), the particle can disappear.

As an example, we have numerically modeled the density of particles
$Q(\widetilde{x},\widetilde{x}^{\prime};\overline{t})$ to the right of the
symmetric barrier ($\Delta=0$) with $\overline{U}=10$ at different
$\widetilde{x}$, when the diffusion starts to the left of the barrier at
$\widetilde{x}^{\prime}<0$. The scaled time we chose $\overline{t}=1\div10$ is
sufficient to reach the spatial domain $\widetilde{x}=1\div3$ starting at
$\widetilde{x}^{\prime}=-3$. The calculated three-dimensional profile of
$Q(\widetilde{x},\widetilde{x}^{\prime};\overline{t})$ is presented in Fig. 5
for the same (as earlier) width of the potential barrier $d=10^{-7}cm$.%

\begin{figure}
[ptb]
\begin{center}
\includegraphics[
height=2.0003in,
width=3in
]%
{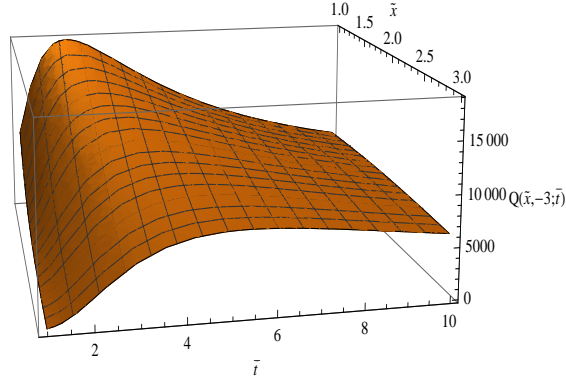}%
\caption{The three-dimensional profile of the particles density $Q(\widetilde
{x},-3;\overline{t})$ to the right of the symmetric barrier. }%
\label{Fig5}%
\end{center}
\end{figure}

One can see the nonmonotonic behavior of the density profile with time
$\overline{t}$ for every fixed $\widetilde{x}$, especially pronounced near the
right barrier boundary (near $\widetilde{x}=1$). This behavior, caused by the
negative sources absorbing the particles (see Eq. (\ref{0e})) and distributed
according to the function (\ref{1}) with $U>0$, $\Delta=0$, is quite different
from the familiar "free" diffusion in the absence of the potential ($U=0$,
$\Delta=0$) which is shown in Fig. 6 for the same parameters as in Fig. 5.%

\begin{figure}
[ptb]
\begin{center}
\includegraphics[
height=2.6481in,
width=3.6737in
]%
{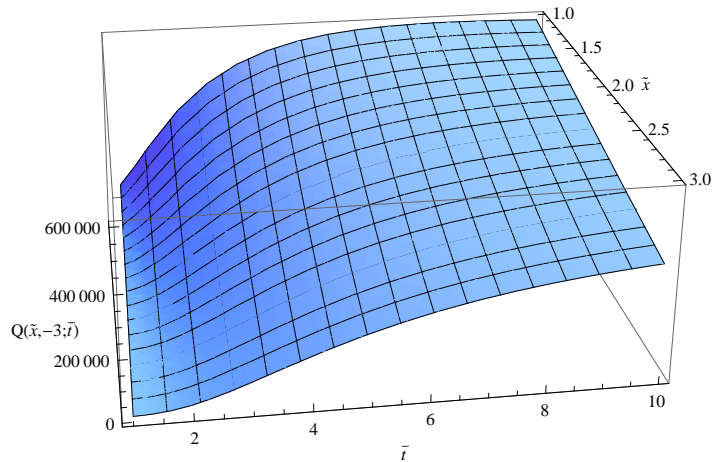}%
\caption{The spacetime density profile $Q(\widetilde{x},-3;\overline{t})$ for
"free" diffusion of the particles. }%
\label{Fig6}%
\end{center}
\end{figure}

\section{Summary}

We have obtained the exact propagator $<x|\exp(-\alpha H)|x^{\prime}>$ ($H$ is
the Hamiltonian for a particle moving in the presence of the asymmetric
rectangular potential) resolving the parabolic-type partial differential
equation. Having obtained the spacetime propagator for the one-dimensional
time-dependent Schr\"{o}dinger equation ($\alpha=it/\hbar$) with a rectangular
well/barrier potential, we at the same time succeeded in finding a propagator
for the Bloch equation ($\alpha=\beta$, $\beta=1/k_{B}t$) for the particle
density matrix and for the diffusion-like equation ($\alpha=t$) by shifting
from real to imaginary time ($t\rightarrow-i\hbar\beta$ and $t\rightarrow-it$,
correspondingly). As an alternative to the conventional path integral approach
to calculating the propagators, we use the multiple-scattering theory for the
calculation of the energy-dependent Green function (a resolvent operator in
(\ref{3b})). The suggested approach is based on the possibility of introducing
the effective potentials (see (\ref{7}) and (\ref{8})) which are responsible
for reflection from and transmission through the potential jumps making up the
rectangular potential (\ref{1}). It provides more of a non-classical picture
of particle scattering at the considered potential as opposed to the
conventional wave point of view.

The solution for the time-dependent Schr\"{o}dinger equation describes the
reflection from and transmission through the asymmetric rectangular potential
as a function of time and thus allows for considering the non-classical
counter-intuitive effects of particle reflection from a potential well and
transmission through a potential barrier in a real situation when a particle
is moving towards the potential (\ref{1}) and then experiencing a scattering
at the potential. These results are also relevant to the fundamental issues of
measuring time in quantum mechanics such as the time-of-arrival (TOA), dwell
time and tunneling time.

The obtained density matrix $\rho(x,x^{\prime};\beta)$ for a particle in a
heat bath and under the influence of the potential (\ref{1}) gives the
probability density (diagonal matrix element) to find a particle in some
spatial point and the quantum correlations (coherences) of different spatial
states $|x>$ and $|x^{\prime}>$ provided by the nondiagonal matrix elements.
The results for the density matrix are numerically visualized, which is
enabled by the fact that they are expressed in terms of integrals of
elementary functions.

The results of the solution of the diffusion-like equation, which can be
interpreted (for the case of a potential barrier, $U>0$) as a diffusion with
the negative sources distributed according to potential (\ref{1}), also have
been numerically evaluated. The corresponding figures demonstrate the
difference between this "diffusion with the holes" and "free" diffusion in the
absence of the potential (\ref{1}).

It is also worth mentioning that all obtained results are also relevant to the
properties of electrons in nanostructures important for spintronics devices
because the potential (\ref{1}) can be used for modeling potential profiles in
such materials.

\end{document}